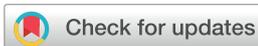

# A toxicology-informed, safer by design approach for the fabrication of transparent electrodes based on silver nanowires

Djadidi Toybou,[ab] Caroline Celle,[a] Catherine Aude-Garcia,[c] Thierry Rabilloud [ID] *[c] and Jean-Pierre Simonato [ID] *[a]

Fabrication of silver nanowires (AgNWs) with fine and independent control of both the diameter (from 30 to 120 nm) and length (from 5 to 120 μm) by concomitant addition of co-nucleants and temperature control is demonstrated, and used for the preparation of size standards. Percolating random networks were fabricated using these standards and their optoelectronic properties were measured and compared with regard to the nanowire dimensions. The transparent electrodes appear suitable for various applications and exhibit excellent performances (*e.g.* 16 ohm sq$^{-1}$ at 93% transparency), with haze values varying from 1.6 to 26.2%. Besides, *in vitro* toxicological studies carried out on murine macrophages with the same size standards revealed that AgNWs are weakly toxic (no toxicity observed below 50 μg mL$^{-1}$ Ag), in particular compared to other silver nanoparticles. Short AgNWs (4 μm) appeared to be slightly more toxic than longer AgNWs (10 and 20 μm). Conversely, long AgNWs (20 μm) induced a more prolonged pro-inflammatory response in murine macrophages. These results contribute, in a safer by design approach, to promoting the use of short AgNWs. The global knowledge dealing with the combination of nanowire dimensions associated with optoelectronic performances and related toxicity should encourage the rational use of AgNWs, and guide the choice of the most adequate AgNW dimensions in an integrated approach.

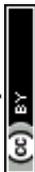

### Environmental significance

The use of silver nanowires is a promising alternative for the fabrication of transparent electrodes. They will be increasingly introduced in consumer devices. The relationship between the nanowires' dimensions and their toxicity is still insufficiently studied. We demonstrate a straightforward method to prepare silver nanowires with controlled dimensions. We propose a global approach, starting from tailor-made nanowires, based on a study of both their physical properties when used as random percolating networks, which is of interest from the technological point of view, and their toxicological behaviour toward murine macrophages, which is of interest from the health, safety & environment point of view. These results contribute, in a "safer by design approach", to promoting the use of short AgNWs.

## Introduction

The recent development of nanomaterials has generated a wave of hope in many fields. The potential of nanomaterials appears boundless, either due to their intrinsic properties (*e.g.* quantum dots and fullerenes), their additional properties for bulk materials (*e.g.* nanocomposites), or when they are used in the form of assemblies (*e.g.* carbon nanotube-based cables). A characteristic and high-potential example of nanomaterial assembly is the fabrication of random networks of metallic nanowires, in particular, silver nanowires (AgNWs).[1–3] Simply put, it is possible to generate random networks of metallic nanowires, above the percolation threshold, which conduct electricity through the nanowire-based metallic lattice and concurrently let the light pass through the empty spaces between the nanowires. This has been widely studied during the last decade and has raised the maturity of this technology up to the industrial level. Many applications using AgNWs have been investigated, including touch screens, light-emitting devices, transparent film heaters and others.[4–12] This system has many advantages such as very low sheet resistance with high transparency, flexibility, and low-cost processing under ambient conditions.

Though many important advances have been realized in this field, some key points remain to be tackled, in particular, toxicity issues for safe industrial use. Since the expected properties are application-dependent and rely on the AgNWs' morphology,[13,14] it would be of great interest to be able to tailor the shape of the nanowires in order to obtain optimal optoelectronic performances for each use. Conventional parametric studies[15,16] as well as other more specific methods[17,18] have demonstrated efficient routes for modifying both the length and diameter, but not independently. An ultrasonic method was also proposed to select the desired length by cutting AgNWs; however, this method leads to poor control over size distribution and generates large amounts of by-products.[19]

[a] *Univ. Grenoble Alpes, LITEN DTNM, CEA, F-38054 Grenoble, France. E-mail: jean-pierre.simonato@cea.fr*
[b] *Univ. Grenoble Alpes, ISTERRE, CNRS, F-38000, Grenoble, France*
[c] *Univ. Grenoble Alpes, Chemistry and Biology of Metals, CEA BIG, CNRS UMR 5249, F-38054 Grenoble, France. E-mail: thierry.rabilloud@cea.fr*

The fabrication of AgNW size standards would also allow toxicological studies to be performed on definite species, in particular to determine whether the two main dimensions (*i.e.* diameter and length) play a critical role in toxicity. This is of tremendous importance since AgNWs will increasingly be introduced in technological devices and consumer products, and thus data are awaited to assess the real toxicity of these nanomaterials. Although several reports have begun to tackle this topic,[20–26] much more remains to be done.

In this article, we propose a straightforward method to modulate independently the diameter and the length of nanowires in a wide range, allowing us to fabricate size standards, *i.e.* AgNW samples with definite lengths and diameters. It has permitted us to perform a direct comparison of the optoelectronic properties of transparent electrodes made with these different nanowires, and to realize comparative experiments on toxicity with macrophages.

## Materials and methods

### Synthesis and purification of AgNWs

In a typical synthesis, AgNO$_3$ (0.68 g) was dissolved in EG (40 mL) at a slow stirring rate in a round flask. In another flask, PVP (average mol. wt 40 000, 1.77 g), NaCl, and the co-nucleant were dissolved in EG (80 mL) at 120 °C. The solution was cooled to room temperature and then slowly added to the first flask within 8 min. The mixture was finally heated at the reaction temperature and cooled down at ambient temperature. The purification of the AgNWs was realized by decantation according to a published procedure.[27]

### Characterization of the mean dimensions of AgNWs

The morphology analysis of AgNWs was performed using a Leo 1530 SEM. Measurement of the diameter and length dimensions was realized with the ImageJ software. The statistical studies were performed on 100 to 200 nanowires for each sample.

### Preparation of AgNW electrodes

The AgNW-based solutions were used for the fabrication of electrodes after dilution in methanol at 0.2 mg mL$^{-1}$ concentration. The silver concentration was measured using an atomic absorption spectrometer (Perkin Elmer AAnalyst 400). The deposition was performed on 2.5 × 2.5 cm$^2$ heated substrates (70 °C) with an automatic SonoTek spray-coater. The performances of the electrodes were measured after cooling, without any post-treatment.

### Performance measurements

Total transmittance and haze values were measured with a UV-visible spectrometer (Varian Cary 5000) equipped with an integrating sphere. The sheet resistance was set as the mean value of at least 5 measurements by using a four-pin probe with a Loresta resistivity meter (EP MCP-T360).

### Toxicological assays

**Nanomaterials.** As control nanomaterials, spherical silver nanoparticles (<100 nm) coated with PVP40 were purchased from Sigma-Aldrich (catalogue number 758329) as a 5 wt% dispersion in ethylene glycol. Working solutions were prepared by dilution in deionized water. The characterization of these nanoparticles has been published previously by some authors of this article.[28] Additional controls consisted of silica-based nanomaterials, either colloidal silica[29] or crystalline silica (reference materials BCR66 and BCR70, purchased from Sigma-Aldrich).

**Cell lines.** The mouse macrophage cell line RAW 264.7 was obtained from the European Cell Culture Collection (Salisbury, UK). The cells were cultured in RPMI 1640 medium + 10% fetal bovine serum. The cells were seeded at 200 000 cells per ml and harvested at 1 000 000 cells per ml. For treatment with nanomaterials, the cells were seeded at 500 000 cells per ml. They were treated with nanomaterials on the following day and harvested after a further 24 hours in culture.

**Neutral red uptake assay.** This assay was performed according to a published protocol.[30] Cells grown in 12-well plates and treated or not with nanomaterials were incubated for 1 h with 40 μg mL$^{-1}$ neutral red (final concentration, added from a 100× stock solution in 50% ethanol–water). At the end of the incubation period, the medium was discarded and the cell layer was rinsed twice with PBS (5 min per rinse). The PBS was removed, and 1 mL of elution solution (50% ethanol and 1% acetic acid in water) was added per well. The dye was eluted for 15 min under constant agitation, and the dye concentration was read spectrophotometrically at 540 nm. The results were expressed as % of the control cells (untreated).

**Cytokine production and persistence experiments.** The tumour necrosis factor (TNF-α), interleukin 6 (IL-6) and monocyte chemoattractant protein-1 (MCP-1) concentrations in the culture supernatants of cells exposed for 24 h to AgNWs and control nanomaterials were measured using the cytometric bead array (CBA) mouse inflammation kit (BD Pharmingen, France) according to the manufacturer's instructions. The measurements were performed on a FACSCalibur flow cytometer and the data were analysed using CellQuest software (Becton Dickinson). Bacterial lipopolysaccharide (LPS) was used as a positive control (200 ng mL$^{-1}$, 24 h). At the end of the exposure period, the medium was removed and saved for cytokine measurements. Fresh complete culture medium (without nanomaterials) was added to the cells and left for 36 h. The medium was then removed and saved, and fresh complete culture medium was added for a final 36 h period. Thus for each nanomaterial, 3 cytokine assays were performed, covering the 24 h exposure period and two post-exposure 36 h time windows.

## Results and discussion

### Controlled synthesis of AgNWs: key factors for independent tuning of length and diameter

The synthesis of AgNWs has been extensively studied since the first report of Xia *et al.*[31] Among different routes to fabricate these nanowires, the most studied process has been the *in situ* reduction of silver salts by the reductive polyol method. This solvothermal process usually includes the use of ethylene glycol as the main solvent, silver nitrate as a cheap silver source, a nucleant with or without co-nucleant(s) to initiate the growth of nanowires and PVP (polyvinylpyrrolidone) as the capping agent which allows orientation of the uniaxial growth.[32] It has already been demonstrated that AgNWs with different lengths and diameters can be obtained "on demand" thanks to the conditions extracted from parametric studies; however, it has not been demonstrated so far that both the diameter and the length could be tuned independently. In other words, modification of the process to adjust one dimension has concomitantly a strong effect on the other dimension (*e.g.* protocol adjustment to modify the mean diameter will strongly impact the mean length of the AgNWs).

It has been previously reported in the literature that the use of a co-nucleant besides NaCl, the molar mass of PVP and synthesis temperature may strongly affect NW morphologies.[33] In particular, halide ions are known to modify the nanowire diameter.[34–36] Bromide ions proved to be efficient in reducing the diameter, but at the same time they also altered the length of AgNWs.[37–39] To begin this study, the influence of $F^-$, $Cl^-$, $Br^-$ and $I^-$ as co-nucleants was investigated and compared to a standard synthesis protocol.[40,41]

The impact of each halide ion on the morphology of as-synthesized AgNWs is presented in Fig. 1. Compared to the standard synthesis without a co-nucleant (Fig. 1a, pink cross, 63 ± 10 nm diameter/8 ± 3 μm length), each added halide ion (molar ratio, 1:1 NaCl:KX) has a noticeable effect on the final morphology of AgNWs. An exception is the case of iodide-modulated synthesis which leads only to irregular nanoparticles. The results obtained with KBr or NaBr as a co-nucleant appeared very close. To confirm this point, NaCl and KCl were also used as nucleants (standard conditions, 0.7 mmol L$^{-1}$), and as shown in Fig. 1(a), AgNWs with similar morphologies (51 ± 7 nm diameter and 6 ± 1 μm length) were produced. This means that using KCl as a co-nucleant in an equimolar ratio with NaCl gives the same result as doubling the amount of NaCl; thus alkali counter-cations do not play a significant role in the reaction mechanism. Nevertheless, it must be noticed that both the diameter and length are affected in the same way when the chloride quantity changes, as already reported elsewhere.[42] Fluoride ions (Fig. 1a – blue triangle) increased the length from 8 ± 3 μm to 12 ± 5 μm, and the diameter to 71 ± 10 nm. Whereas the effects of fluoride and iodide ions have not been reported extensively, diameter shrinkage with the use of a bromide-based co-nucleant is in fair agreement with the literature.[37–39,43]

To further investigate the effect of bromide as a co-nucleant, we performed experiments with varying concentrations of KBr, from 0 to 0.7 mmol L$^{-1}$. The results are presented in Fig. 1b. It appears that the diameter decreases from 63 ± 10 nm down to 20 ± 3 nm with increasing amount of bromide ions, while the length increases from 6 ± 3 to 12 ± 5 μm. Nanowires with 15 nm diameter can also be obtained at higher concentrations of bromide, but the amount of metallic by-products becomes excessive. As shown here, and in agreement with literature reports, adjustment of the KBr

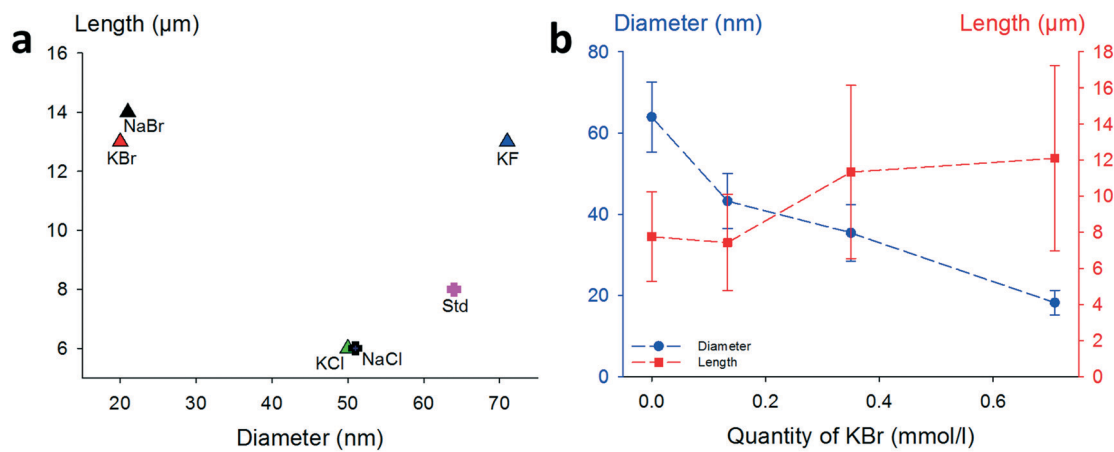

Fig. 1 Influence of several halide ions on the morphology of AgNWs. (a) Lengths of AgNWs synthesized by a standard protocol[41] (pink dot, NaCl as the only nucleant) and modified procedures with additional halide salts as co-nucleants (KBr, NaBr, KCl, NaCl, and KF). (b) Effect of KBr concentration on diameters and lengths of AgNWs.

quantity leads to a fine control of the diameters but changes significantly the lengths.

Thus, in our goal to fabricate "on demand" AgNW size standards with defined lengths and diameters, we carried out experiments by setting the KBr amount at 350 mmol L$^{-1}$ to control the diameter, and we introduced NaCl in various amounts in order to modulate the length. The results plotted in Fig. 2a show that when the chloride concentration was varied between 0.5 and 2.5 mmol L$^{-1}$, we obtained a collection of AgNWs with lengths ranging from 28 ± 9 down to 4 ± 2 μm. For higher concentrations of chloride, nanoparticles were obtained as the sole product. These results demonstrate that by adequate choice of the nature of co-nucleants and by fine adjustment of their concentrations, the variation of length is actually possible while keeping the diameter almost unchanged (Fig. 2b). This is a straightforward route to select the desired length at constant diameter. The mean length modification can be ascribed to the increased (or decreased) number of seeds, which depends on the molar ratio between chlorine and silver. This is consistent with the report from Buhro and colleague[44] who demonstrated, in the early stage of the polyol synthesis, the impact of chloride concentration on the generation of nucleation sites.

Whereas diameters below 50 nm are desirable for most optoelectronic applications requiring a low haze value, in some specific uses like photovoltaics, larger diameters are expected.[5] In our effort to modulate length and diameter independently, we looked at a protocol to modify the diameter while keeping the length mostly unchanged. It was previously demonstrated by Unalan and colleagues that polyol synthesis carried out at low temperatures generates thick wires (with diameters higher than 300 nm) and that temperature plays a crucial role in the nanowire formation.[15] We used our standard protocol at various temperatures. The results are presented in Fig. 2c and d. We observed that the mean diameter decreased drastically, from 90 ± 10 nm down to 50 ± 8 nm, when the temperature was raised from 150 to 180 °C. At the same time, the mean length of the AgNWs remained almost constant. This demonstrates that it is also possible to change the diameter of AgNWs while keeping the length mostly unchanged by selecting the adequate reaction temperature.

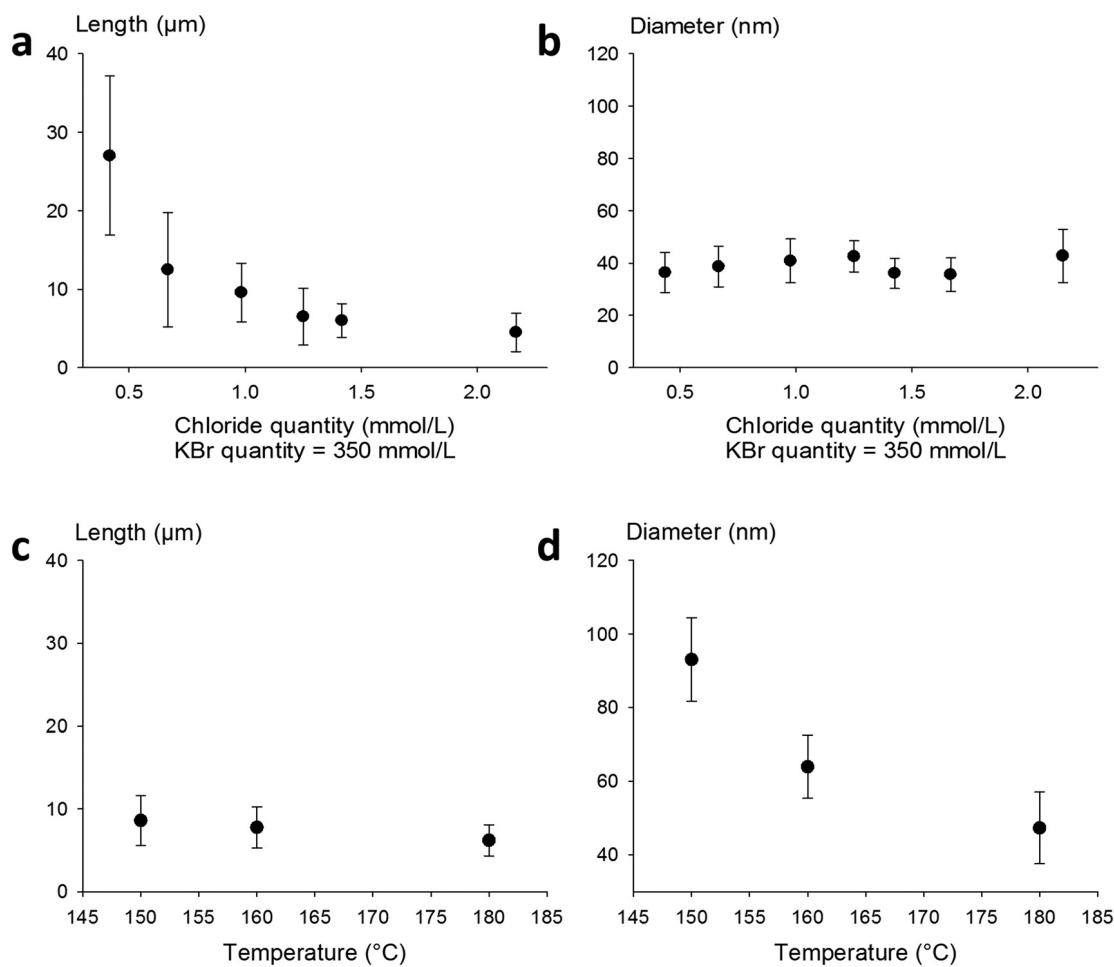

Fig. 2 Tuning of the length of thin AgNWs by NaCl concentration and reaction temperature. (a) Length as a function of the chloride concentration, with a constant KBr concentration of 350 mmol L$^{-1}$. Increasing the quantity of chloride decreases the mean length of AgNWs. (b) The diameter of AgNWs remains constant around 40 nm while the length decreases from 28 μm to 3 μm. The KBr concentration is set at 350 mmol L$^{-1}$. (c) The length remains almost constant while the reaction temperature is modified. (d) Diameter tuning by reaction temperature modification, where increasing the temperature induces a diameter decrease.

**Fabrication of AgNW size standards.** Thanks to the protocols presented hereinbefore, we were able to prepare different AgNW size standards, as shown in Fig. 3. Combinations of any lengths or diameters in the range of 5–120 μm and 30–120 nm, respectively, were achievable. This gives access to calibrated samples of nanowires. Histograms showing the distribution in diameters of three different samples of 10 μm long AgNWs with various mean diameters are presented in Fig. 3a, and histograms of three different populations with different lengths of 40 nm diameter AgNWs are shown in Fig. 3b. In each case, the overlap between two adjacent statistical populations was calculated to be below 15%. SEM images are also presented to illustrate the major differences obtained between the different samples in Fig. 3c; they show undoubtedly how both diameters and lengths can be tuned according to specific synthetic protocols. The main criteria to tune the synthesis of the AgNW standards are schematized in Fig. 3d. We also verified that the proposed conditions are scalable. We performed experiments using up to 5 L semi-batch reactors, which allowed us to confirm that the synthesis of AgNWs with pre-determined dimensions can be carried out "on demand".

**Optoelectrical performances of transparent electrodes: impact of AgNWs' dimensions on haziness.** When assembled in the form of random networks above the percolation threshold, it is known that these materials can provide a relevant alternative to TCOs (transparent conductive oxides) for the fabrication of transparent conductive materials. In this case, the properties of interest (*e.g.* transparency and sheet resistance) are measured at the macroscopic scale, but depend strongly on the nanosized building blocks. We thus fabricated transparent electrodes using various AgNW standards with aspect ratios (*i.e.* length/diameter ratios) ranging from 55 to 1000. For the sake of clarity, the different AgNW mean dimensions are expressed as DaLb, with a and b indicating the mean diameter in nm and the mean length in μm, respectively. The transparency, conductivity and haze factor (*i.e.* the diffused part of the transmitted light) of the electrodes were measured and compared.

A typical plot of the transmittance as a function of the sheet resistance is presented in Fig. 4a. This series of dots was measured using D30L10 AgNWs. As expected, for these dimensions and for highly purified nanowires, excellent optoelectrical performances were obtained, with for instance 97% transparency for a 50 ohm sq$^{-1}$ sheet resistance. This compares very well with the state-of-the-art for metallic nanowire-based technology, and also with ITO or any other kind of transparent electrode. We chose to limit the field of study to nanowires with at most 20 μm length because longer AgNWs are much more difficult to synthesize, to handle, to formulate and to deposit with high homogeneity on a large substrate area using low-cost solution deposition techniques. In Fig. 4b, the haze values are presented as a function of the total transmittance for AgNWs with different diameters and for different lengths (5, 10 and 20 μm long). This graph shows a linear relationship between the haze value and the total transmittance regardless of the length. The dimension effect is very important since the haze values vary from 16% down to less than 2%. Very low haziness, compatible with optoelectronic applications, is demonstrated for small diameters, which is consistent with previous studies,[45–47] and very interestingly, it is also shown that even with 90 nm diameter AgNWs, it is possible to achieve low haze values for long nanowires. The haze factor reported in the literature for sub-15 nm diameter silver nanowires reached very low values below 1% at 94.5% transmittance.[48] However, reducing the diameter of AgNWs below the mean free path of electrons (~40 nm) is risky due to charge carrier scattering at the interface and to material instability under operational stress. The histogram in Fig. 4c shows haze values at given sheet resistances for AgNWs with various lengths and diameters. When the densities of AgNWs are decreased (i.e. increasing sheet resistances), the haze value is lowered drastically regardless of the size standards. The impact of the length is obvious since 20 μm long nanowires are much less light diffusive than 10 μm long nanospecies. For instance, in the case of 40 nm diameter AgNWs, the haze value is more than doubled when the length is reduced from 20 μm down to 10 μm. This is expected by percolation theory. Indeed, for finite size scaling in stick percolation dealing with large size systems, the critical number density ($N_c$) of sticks required for percolation is given by $N_c \times L^2 = 5.637$ where $L$ is the length of nanowires.[49] This indicates that for a two-fold increase of the length of the AgNWs, the required $N_c$ to reach percolation is divided by four.[17,50] As shown by De et al. and by Lagrange et al., percolation theory dominates the electrical behaviour of networks associated with low density.[51,52] However, it should be emphasized that in our examples, the electrodes are fabricated with densities far above $N_c$, but the length effect is still clearly perceptible. The diameter effect is also visible for both 10 μm and 20 μm long AgNWs, and as expected, small diameters lower the haziness. Pictures of two electrodes are included in Fig. 4d to illustrate the pronounced optical difference (transparency and haziness) for different densities of AgNWs.

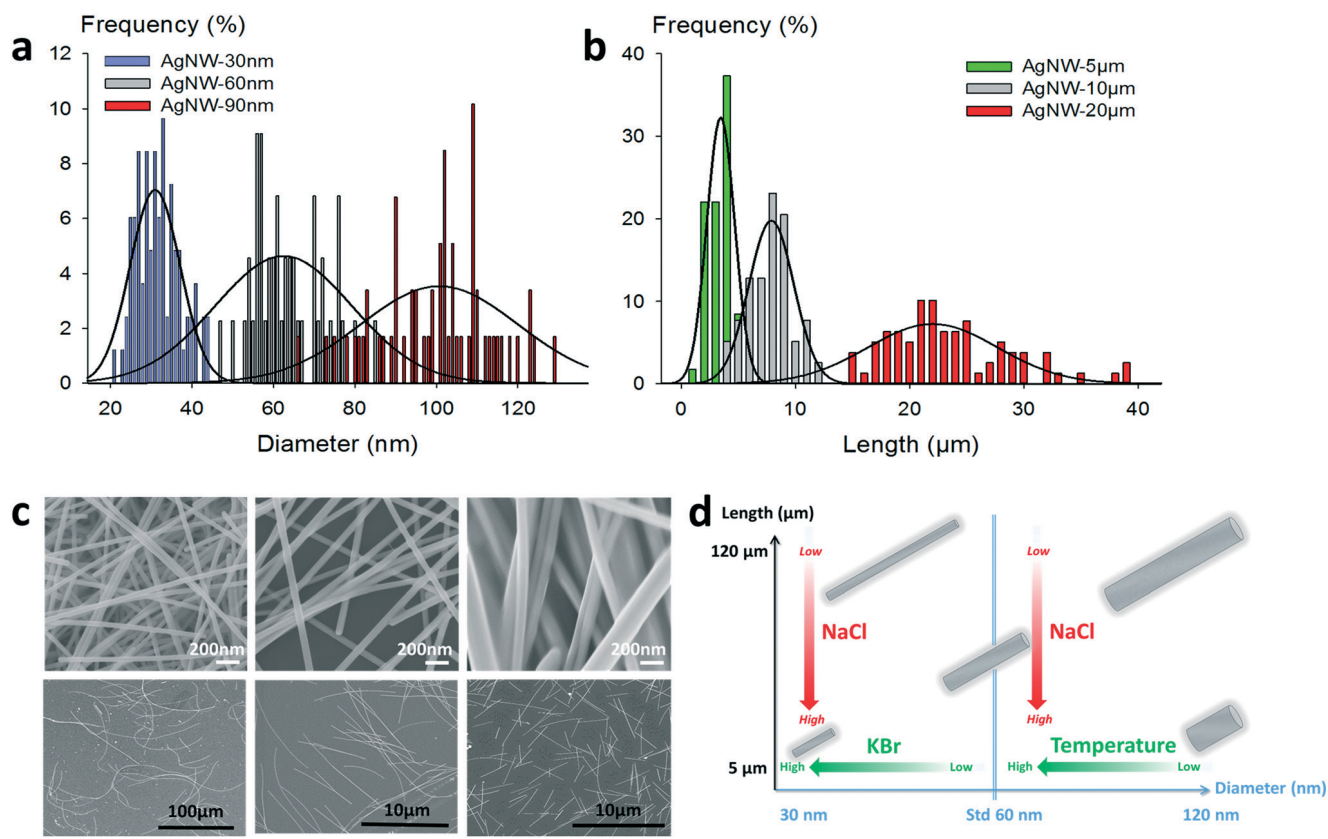

Fig. 3 Independent tuning of the diameter and length of silver nanowires. (a) Histograms of three different mean diameters for 10 μm long AgNWs. (b) Histograms of three different mean lengths for 40 nm diameter AgNWs (c) SEM pictures of various lengths and diameters. Top, from left to right: mean diameters of 30 nm, 60 nm, and >100 nm; bottom, from left to right: mean lengths of >100 μm, 10 μm, and 5 μm. (d) Schematic diagram to control the morphology of AgNWs.

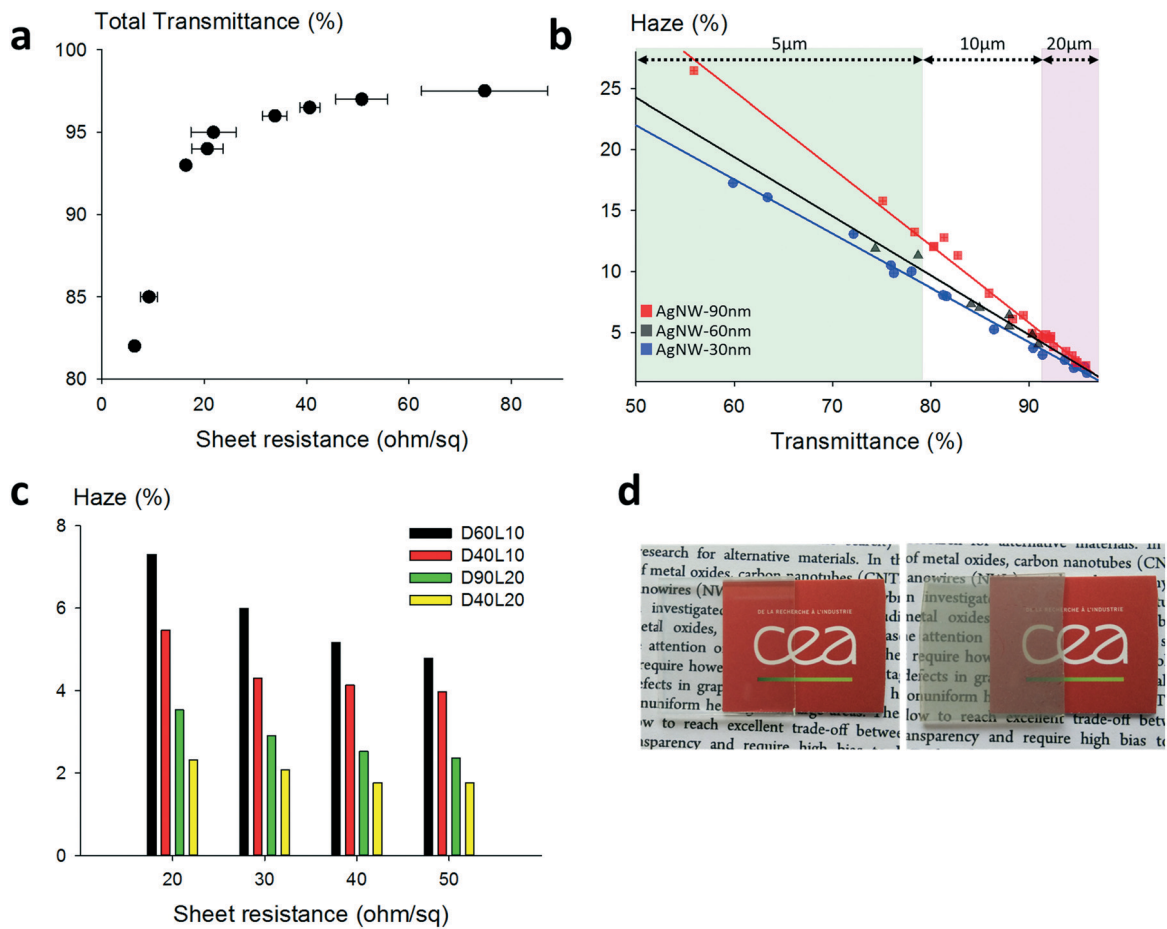

Fig. 4 Optoelectrical properties of transparent electrodes fabricated with various AgNW size standards. (a) Total transmittance as a function of the sheet resistance for D30L10 AgNWs. (b) Haze values for 30, 60 and 90 nm diameter AgNWs as a function of the total transmittance. The green part represents measurements made on 5 μm long AgNWs, the white part for 10 μm length and the pink part for 20 μm length. The solid lines correspond to linear regressions for each diameter. (c) Comparison of the haze level on given sheet resistances for standard AgNWs (D60L10), thin and short (D40L10), thick and long (D90L20) and thin and long nanowires (D40L20). (d) Left, a 96% transparent electrode with a very low haze value (1.6%); right, a 60% transparent electrode with high haziness (13.0%).

**Toxicity of AgNW size standards.** Even if silver toxicology is well known,[53,54] the toxicity of silver-based nanomaterials is difficult to predict because of their different shapes, which can contribute to different biological responses regarding the plurality of microorganisms. It is described in the literature that low-aspect ratio nanosilver species such as nanoparticles exhibit toxicity mainly relying on silver ion release, either extracellularly (for bacteria) or due to internalization and dissolution in the lysosomes for animal cells.[28,55,56] For high aspect ratio silver nanomaterials, like AgNWs, the toxicity effect may be driven by dissolution on the one hand; on the other hand, the form factor may contribute largely, as AgNWs exhibit a fibre shape having similarities with asbestos. Different toxicity approaches are currently ongoing to decipher possible mechanisms in order to avoid toxicity caused by the aspect ratio, and some research studies are carried out to establish safety thresholds like for asbestos.[57–59]

In this study, we investigated the basic responses of murine macrophages in the presence of AgNWs. We chose macrophages because they are in charge of removing nanomaterials, and because they are also the main actors at play in insoluble fibre toxicity. Thanks to the fabrication of AgNW size standards described above, toxicological experiments were performed on macrophages to discriminate the effects of the diameter and length.

Previous studies dealing with the potential toxic effect of AgNWs on lung and macrophage models revealed a length-dependent toxicity for both *in vitro* and *in vivo* studies, with short nanowires causing less inflammatory responses.[20,57,60–62] These different studies compared various lengths; however, a possible diameter effect was not directly tackled by comparative experiments, or both diameters and lengths were changed at the same time.

Prior to toxicity experiments on macrophages, we evaluated the contribution of dissolved species. The dissolution of relevant morphologies of nanowires was monitored over a period of 24 h in medium only under working conditions. The AgNWs were compared to silver nanoparticles coated with PVP with a mean diameter close to standard AgNWs (≈60 nm) and citrate-coated silver nanoparticles which are known to dissolve quickly. The results are shown in Fig. 5a.

Regardless of their dimensions, AgNWs exhibit similar dissolution rates. Silver nanoparticles coated with PVP dissolved slightly faster than the nanowires. For all the AgNWs, the solubilized part of silver was found to be low (<10%), even after 24 h. The corresponding $Ag^+$ concentration was lower than the known silver ion cytotoxicity (1 μg mL$^{-1}$); hence subsequent toxicity measurements on macrophages were considered relevant to evaluate the potential intrinsic toxicity of AgNWs.

**AgNWs show low and length-dependent toxicity for macrophages.** Cytotoxicity assays were performed on RAW264.7 murine cells, and the neutral red uptake assay was used to determine the potential adverse effects for each morphology of AgNWs. Various morphologies were chosen to check independently the potential diameter or length effects. For instance, we used AgNWs with the diameter dimension set at 40 nm with various lengths (4, 10 and 20 μm), and 10 μm long

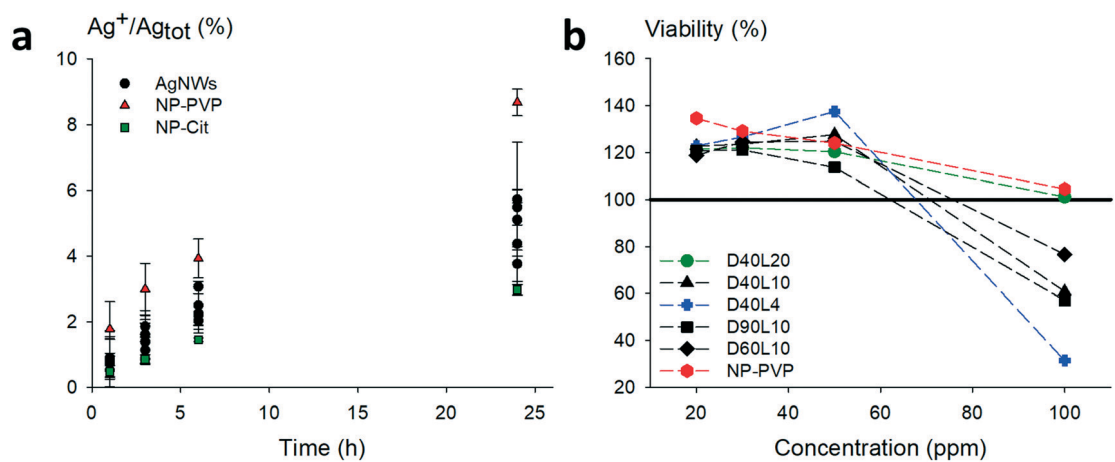

Fig. 5 Dissolution rate of the studied nanomaterials and cell viability of RAW264.7 murine cells treated with the nanomaterials. (a) Dissolution monitored for 24 h at 37 °C in RPMI medium containing only the nanomaterials at 2.5 μg mL$^{-1}$ concentration except for less concentrated citrate-coated nanoparticles (0.4 μg mL$^{-1}$). PVP-coated nanoparticles (NP-PVP, red triangle) dissolve faster than nanowires (AgNWs, black dot). The total dissolution of all the nanomaterials is less than 10% in the medium. (b) The cells were incubated for 24 h with different concentrations of the nanomaterials (from 20 to 100 ppm). The shortest nanowires (D40L4) exhibit a higher toxicity, AgNWs with 10 μm length (D40L10, D60L10, and D90L10) exhibit a similar toxicity regardless of the diameter dimension, and long nanowires with 20 μm length (D40L20) and PVP-coated nanoparticles (NP-PVP) show a very low toxicity. The LC50 of the short nanowires, which exhibit the highest toxicity, is higher than 80 ppm.

AgNWs with various diameters (40, 60 and 90 nm). No decrease of cellular activity up to high doses (50 ppm nanomaterials) was observed, as evidenced in Fig. 5b. The shortest AgNWs (length of 4 μm) presented the lowest IC50 at 100 ppm concentration, followed by the three other nanowires with 10 μm length. Longer AgNWs (20 μm length) did not present a cellular activity decrease under our experimental conditions. This length effect can be related to the uptake ability of macrophages. Phagocytosis of short nanowires (4 μm) can be easily completed, leading to silver ion release inside the cells, as with the well-described silver nanoparticle toxicity. The 10 μm long nanowires, with different diameters, show a similar inhibition behaviour. The quasi non-toxicity of long AgNWs (20 μm length) can probably be ascribed to the fact that the nanowires were not internalized, as the threshold length of phagocytosis was exceeded. These results suggest that AgNWs induce cytotoxicity in a length-type-dependent manner, which is consistent with a reported study on RAW264.7 treated with carbon nanotubes of different lengths.[63] These results are in accordance with the literature, with cytotoxicity being correlated with the degree of cell uptake and the amount of ionic silver due to intracellular dissolution.[62,64]

**Long AgNWs induce a slight but persistent pro-inflammatory profile.** Macrophages are known for their capacity of phagocytizing particulate substances, such as pathogenic agents but also particulate chemicals (*e.g.* oxidized lipoprotein particles in the case of atherosclerotic foam cells). They also play a strong role in immune reaction modulation through their ability to release various signalling molecules such as pro- or anti-inflammatory cytokines, depending on the nature of the phagocytized particles. They also trigger these mechanisms upon frustrated phagocytosis (*e.g.* in the asbestos case), which makes them an attractive choice to study possible pro-inflammatory responses. To evaluate this response, we measured two major inflammatory cytokines: interleukin 6 (IL-6) and tumour necrosis factor alpha (TNF-α) at the highest non-toxic concentration (50 ppm). The inflammatory responses due to the nanowires were compared, on the one hand, to that of amorphous and crystalline silica particles (LS30, BCR66, and BCR70) known to induce a low but well-documented inflammatory response[65–68] and on the other hand, to bacterial lipopolysaccharide (LPS) as a positive control known to induce very strong pro-inflammatory responses.

Fig. 6a shows the relative amount of secreted TNF-α by macrophages in response to the different nanowires. After 24 h, we can notice a similar production rate for all the morphologies of AgNWs. However, D40L20 is the only AgNW that induces a slightly higher TNF-α production than silica. For IL-6 (Fig. 6b), the production induced by AgNWs was very low and always lower than those induced by silica. Overall, the bar charts for IL-6 and TNF-α show a globally low secretion for each cytokine after a 24 h exposure to nanomaterials, representing less than 1% of the LPS-induced production. These results show that the AgNWs induce a weak increase in cytokine release, which correlated positively with the length of the wires, as expected for fibre-like materials.

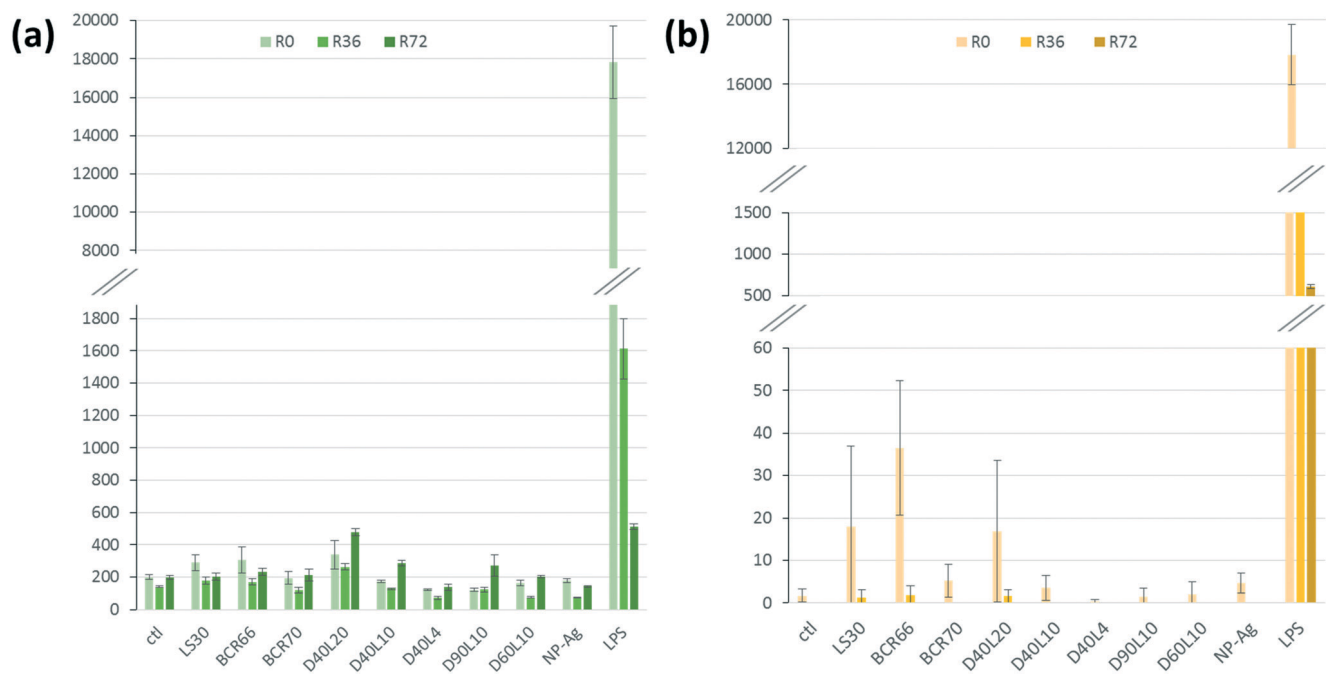

**Fig. 6** Pro-inflammatory cytokine secretion after treatment with nanomaterials (R0) and persistence results at 36 h (R36) and 72 h (R72). The secretion level was compared to LPS as a positive control and silica-based nanomaterials (LS30, BCR66, and BCR70) as the nanomaterial control. (a) TNF-α secretion is very low (the same level as the control) for all AgNWs, and lower than the silica control (BCR66), except for D40L20. (b) The IL-6 secretion of cells treated with AgNWs is similar regardless of the length or diameter, and at the same level as the negative control. For both IL-6 secretion and TNF-α secretion, the comparative secretion with LPS is very low (<1%).

One of the major mechanisms at play in persistent fibre/particle toxicity is the presence of a prolonged pro-inflammatory response over time. In order to test this possibility by our *in vitro* system, we studied the evolution of the production of TNF-α and IL-6 over time after a 24 h exposure to nanomaterials (Fig. 6). LPS-induced TNF-α release was both acute and very transient. Compared to the control, none of the short nanowires, the amorphous silica or the silver nanoparticles induced any significantly higher TNF-α release. Only the small crystalline silica (BCR66) and the long nanowires (20 μm) induced a weak but sustained TNF-α release. Although the shape of the curves is different for IL-6, with a prolonged LPS-induced release, the same trend can be observed, with only the small crystalline silica (BCR66) and the long nanowires inducing a prolonged IL-6 release compared to the control. These results suggest that the long nanowires have an inflammatory effect that is sustained over time. To the best of our knowledge, this is the first reported study on the persistence of the inflammatory response *in vitro*.

Our results are in line with those obtained using *in vivo* models[20,26,60] and suggest that simple *in vitro* models can be predictive of the inflammatory potential of nanomaterials. In addition, these *in vitro* systems are much cheaper and ethically more acceptable than *in vivo* experiments.

Taken collectively, all these data allow us to propose a "safer by design" approach for the use of AgNWs, depending on the applications. Different performances for both electrical and optical aspects (including haziness) can be achieved, depending on the density and on the dimensions of AgNWs. According to the toxicological results, a rational approach would be to foster the use of nanowires up to 10 μm length since they can combine high optoelectrical performances and minimal biological effects.

## Conclusions

By careful modifications of the polyol process, in particular, precisely defined halide ion concentrations and temperature setting, independent fine tuning of the diameter and length of AgNWs has been successfully demonstrated. Size standards were prepared accordingly, and used for the fabrication of transparent electrodes. Excellent optoelectrical performances were obtained and the haze value was varied from 1.6 to 26.2%. In the meantime, we performed *in vitro* toxicological assays on murine macrophages. The AgNWs were found to be weakly toxic for macrophages and showed a length-dependent toxicity, with the toxicity decreasing with length. Conversely, the cytokine release assays showed a weak but significant pro-inflammatory potential for long nanowires (20 μm), and the observed persistence of the response, although at weak levels, calls for attention when devising the rules for the use of AgNWs in consumer products and in the recycling of such products. This study set out to take a holistic view of AgNWs, with a clear relationship between AgNW dimensions and both their optoelectronic performances, when used in the form of random percolating networks, and their toxicity and pro-inflammatory potential. Since AgNWs will certainly be increasingly introduced in commercial devices in the short term, it seems important to have a global vision of these nanomaterials, which will ineluctably encourage the rational use of AgNWs in a "safer by design" approach.

## Conflicts of interest

There are no conflicts of interest to declare.

## Acknowledgements

This work was supported by the Labex Serenade (no. ANR-11-LABX-0064) funded by the "Investissements d'Avenir" French Government program of the French National Research Agency (ANR) through the AMIDEX project (no. ANR-11-IDEX-0001-02). In particular, the authors thank Pr L. Charlet for coordinating the "nanosilver topic", and they are grateful for the funding of D. Toybou's PhD grant.

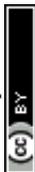